\documentclass[preprint, 1p]{elsarticle}
\usepackage{graphicx}
\usepackage{amsmath}
\journal{Computer and Electrical Engineering}

\begin{document}

\begin{frontmatter}
\title{On the Performance of Adaptive Modulation in Cognitive Radio Networks}
\author[diuoa]{Fotis~Foukalas\corref{cor1}}
\cortext[cor1]{Corresponding author}  
\ead{foukalas@di.uoa.gr}
\author[teil]{~George~T.~Karetsos}
\ead{karetsos@teilar.gr}
\address[diuoa]{Department
of of Informatics and Telecommunications, National Kapodistrian University of Athens, Ilisia, Athens, Greece}
\address[teil]{Department of of Information Technology and Telecommunications, TEI of Larissa, Larissa, Greece}

\begin{abstract}
We study the performance of cognitive radio networks (CRNs) when incorporating adaptive modulation at the physical layer. Three types of CRNs are considered, namely opportunistic spectrum access (OSA), spectrum sharing (SS) and sensing-based SS. We obtain closed-form expressions for the average spectral efficiency achieved at the secondary network and the optimal power allocation for both continuous and discrete rate types of adaptive modulation assuming perfect channel state information. The obtained numerical results show the achievable performance gain in terms of average spectral efficiency and the impact on power allocation when adaptive modulation is implemented at the physical layer that is due to the effect of the cut-off level that is determined from the received signal-to-noise ratio for each CRN type. The performance assessment is taking place for different target bit error rate values and fading regions, thereby providing useful performance insights for various possible implementations. 
\end{abstract}

\end{frontmatter}

\section{Introduction}
Cognitive radio networks (CRNs) can enhance spectrum utilization of licensed wireless networks, also known as primary networks (PN), when certain conditions apply \cite{Hay05} \cite{Khal10}. The underutilized or unused spectrum resources can be exploited by the so called cognitive or secondary networks (SNs) as long as their operation is not harmful for the PN. CRNs are categorized either as opportunistic spectrum access (OSA) or as spectrum sharing (SS) ones \cite{Zhao07} \cite{WangB11}. OSA CRNs rely on exploiting spectrum bands when they are not being used by the PN and the power level is adequate for transmission. SS CRNs rely on the co-ordinated sharing of a spectrum band among the PN and the SN. Furthermore if an SS CRN employs spectrum sensing, then it is called sensing-based spectrum sharing CRN \cite{Kang09}. All three CRN types exploit channel state information (CSI) in order to provide enhanced spectral efficiency over the considered wireless channels \cite{Mus09}\cite{Had08}\cite{WangH10}. To this end, one of the main techniques employed is power control which regulates the transmission of the SN users while protecting the PN ones by exploiting the CSI \cite{Xin09} \cite{Gold97b}. 

On the other hand, adaptive modulation is one of the key techniques used in most broadband wireless communication standards such as the high speed packet access (HSPA) and the worldwide interoperability for microwave access (WiMAX), since it allows for the optimal utilization of fading channels \cite{Al00} \cite{3GPP36201} \cite{IEEE80216}. In particular the transmission rate is adapted based on information regarding the channel conditions that are estimated at the receiver's side and made available at the transmitter throughout a feedback channel \cite{Salam10}. When adaptive modulation is implemented in conjunction with power control at the physical layer, then a variable rate variable power (VRVP) modulation is considered  \cite{Gold97} \cite{Gold05}. Two alternatives of VRVP have been specified known as continuous rate (CR) and discrete rate (DR). The latter is more practical from implementation point of view since the continuous fashion can not be easily accomplished. 

In this work we investigate the incorporation of adaptive modulation in OSA, SS and sensing-based SS CRNs in order to assess the achievable performance in terms of spectral efficiency and to derive the optimal power allocation for each implementation. For adaptive modulation we consider several constellation numbers and bit error rate (BER) target values. Closed form expressions regarding the average spectral efficiency over fading channels and the corresponding required optimal power allocation that maximizes the average spectral efficiency are derived. A Rayleigh fading channel model is considered for characterizing the fading effects on digital transmissions in urban wireless mobile communications \cite{Skla97}. The obtained numerical results highlight the performance of adaptive modulation for each CRN type and the corresponding achievable gain in average spectral efficiency. Furthermore, we derive and evaluate the results of optimal power allocation that indicate the additional transmit power demands. The proposed approach is practical and essential for fourth generation (4G) cellular networks and beyond, which already adopt adaptive modulation and in the near future will adopt CR techniques such as the recently introduced carrier aggregation technique \cite{Iwa10} \cite{3GPP36912}.  

To the best of our knowledge, such a comprehensive study has not been carried out so far. The authors in \cite{Li08} consider a SS CRN with one primary user (PU) and multiple secondary users (SUs) and propose a joint adaptive modulation and power allocation while taking into account the protection of the PU and the quality of service (QoS) of the SUs. The main objective of this work is to minimize the total power consumption while keeping both the received interference level at the PU and the target values of the received signal-to-interference-noise-ratio (SINR) at the SUs at an acceptable level. In this study the received SINR represented the QoS metric used for the SUs. The authors in \cite{Shil08} proposed a  joint bandwidth allocation, adaptive modulation and power control in a space time block coding system in order to maximize the spectral efficiency and the total throughput under several bandwidth constraints. In \cite{Gao08}, adaptive modulation is considered in a distributed ad-hoc CRN where energy efficiency is the performance metric. In \cite{Chen11}, authors analyse the performance of adaptive modulation over Nakagami-m fading channels for opportunistic access without considering the SS cognitive paradigm. In addition the IEEE 802.22 working group has not yet considered the incorporation of adaptive modulation in CRNs and its possible practical deployment \cite{IEEE80222} \cite{Cord05}. 

The reminder of this paper is organized as follows. In Section II we provide the system models employed when adaptive modulation is implemented in OSA, SS and sensing-based SS CRNs respectively. In section III the performance analysis of the proposed schemes is taking place. In section IV, we  provide numerical results that illustrate the behaviour of the assumed deployments. Finally a summary of this work is presented in section V.  
  
\section{System models}
In this section we provide the system models that were utilized for investigating the incorporation of adaptive modulation in the three CRN types.

\subsection{Opportunistic Spectrum Access CRNs}
We assume an OSA CRN with $c\in C$ channels and $u\in U$ users where each
user $u$ is served relying on a spectrum pooling strategy that first
serves the PU and subsequently the SUs \cite{Had08}. In this sense a set of channels is shared among the PU and the SUs. Thus the SU cares only about the availability of channels that are not used by the PU and does not consider any interference related aspects. Fig.1 shows the system model of the considered OSA CRN. The model shows one PU and one SU and the channels $C$ which are available for access from both user types at time $i$  and $i+1$ respectively. For instance, at time $i$ the PU occupies two channels in which the specified level in SNR has been achieved and in the same way the SU occupies two different channels in the considered spectrum at time $i+1$.  The channels are assumed in a fading environment with independent and identically distributed (i.i.d.) channel gain $\sqrt{g(i)}$ and additive white Gaussian noise (AWGN) $n(i)$, both at time $i$.

\begin{figure}[p]
\centering
\includegraphics[width=4.5in]{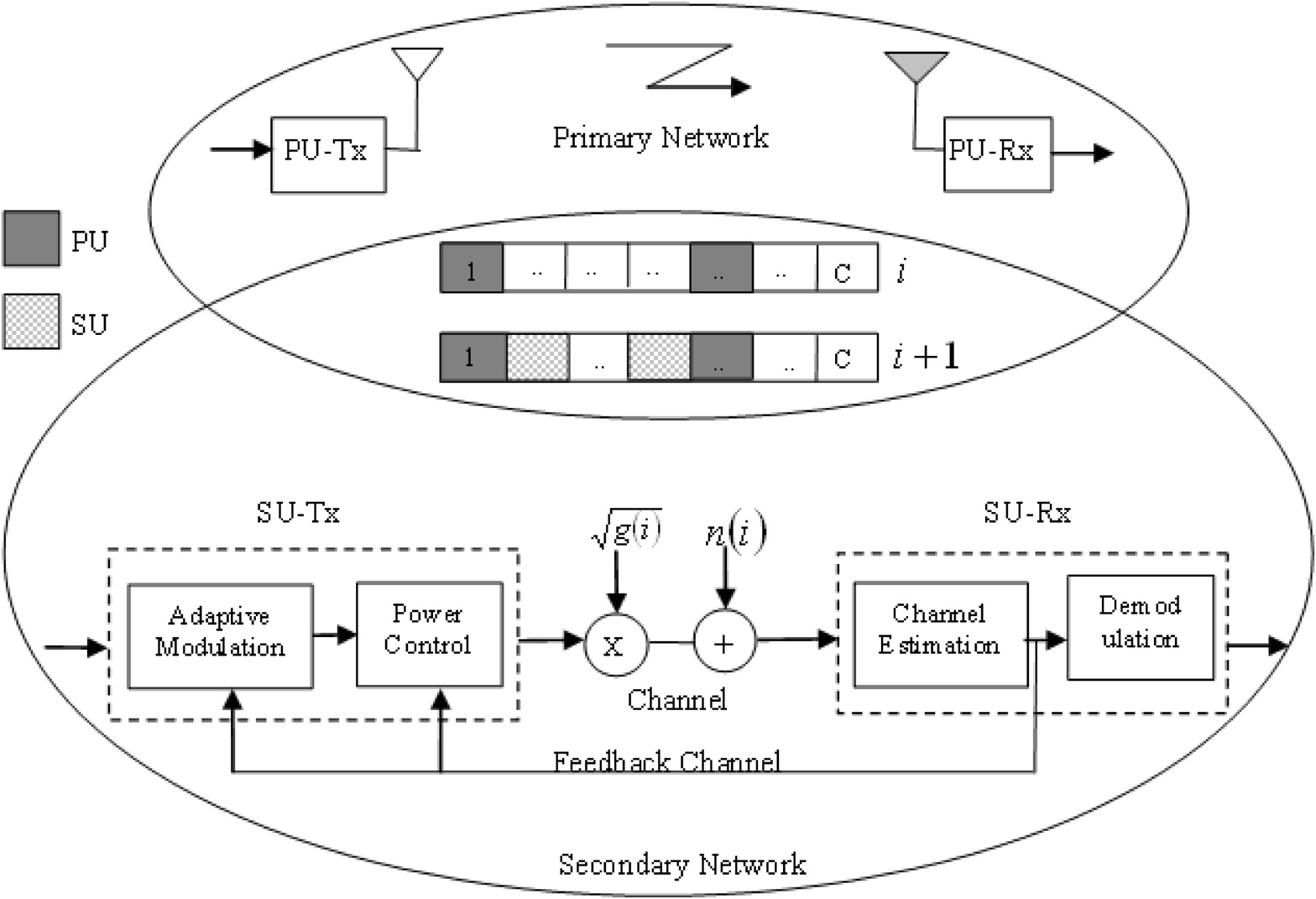}
\caption{System model of Opportunistic Spectrum Access CRN}
\label{fig_syst_mod}
\end{figure}

 The average transmit power over the fading channel is $\overline P$ and the AWGN is with power density $N_0/2$. A SU can access an idle channel $c$ if and only if a predefined level on the instantaneous transmit power $ P $, is achieved. This level is determined from the CSI which represents the received
SNR, $\gamma$ that is equal to $g(i)P/N_0$   at time $i$. Thus, the transmit power $P$ is controlled by $\gamma$ and we denote it as $P(\gamma)$. This policy is known as optimal power allocation and relies on the channel state estimation $\hat g(i)$ \cite{Al99}. We assume that the estimation is performed at the receiver using a training-based channel estimation \cite{Mey98}. The channel estimation $\gamma$ is also available at the transmitter side via a feedback channel \cite{Gold05}. We assume that the CSI is perfectly available at the receivers i.e. PU-Rx, SU-Rx and that the feedback channel does not induce any delays on the CSI's transmission. This can be achieved for example in cellular networks where power control is used. 

The system first determines if a user $u$ can access a channel $c$ and in the sequel it selects the appropriate M-ary Quadrature Amplitude Modulation (M-QAM) from the set $M$ according to the estimated CSI \cite{Al00}. Furthermore, we make the following assumptions for the considered system: a)the total system's bandwidth is $ B $ and is divided into $ C $ channels b) the transmission of each symbol is accomplished with a symbol period $T_s=1/B$ using ideal raised cosine pulses and c) the fading channel is varying slowly in time, i.e. the receiver is able to sense and track the channel fluctuations. Such a channel corresponds to a block flat fading channel model with an average received SNR, $\overline\gamma$ \cite{Big01}.  

\subsection{Spectrum Sharing CRN}
We consider now a SS CRN model with one PN and one SN as depicted in Fig.2. In SS CRNs, the primary user (PU) and the secondary user (SU) share the same channel band for data transmission \cite{Mus09}. All the links are assumed as fading channels with AWGN. The primary link consists of a PU transmitter (PU-TX) and a PU receiver (PU-RX) with channel power gain $g_{pp}$, while the secondary link consists of a SU transmitter (SU-TX) and a SU receiver (SU-RX) with channel power gain $g_{ss}$. Furthermore, the interference links are assumed from PU-TX to SU-RX with channel power gain $g_{ps}$ and from SU-TX to PU-RX with channel power gain $g_{sp}$. The AWGN is assumed to be an independent random variable with mean zero and variance, $N[0,\sigma^2]$. The average and the instantaneous transmit power of the SU-Tx over the fading channel is $\overline P$  and  $P$ respectively. Thus, the CSI which  represents the received SNR $\gamma_{ss}$ at the secondary link is equal to $g_{ss}P/N_0B$, assuming that the interference from the PU is negligible. CSI is perfectly available at the secondary receivers and the secondary link employs power control. A feedback channel makes available the received SNR at the secondary link $\gamma_{ss}$ to the SU-Tx without any delay. The power control mechanism at the SU-Tx and the corresponding optimal power allocation has to protect the PU’s transmission in a SS CRN where the channel is shared with the SU. Efficient power allocation policies can lead to the maximization of the overall achievable throughput and are of primary importance for CRNs \cite{Zay08}. Details regarding the approaches adopted for power control and power allocation in SS CRNs are provided in section III.  

\begin{figure}[tpb]
\centering
\includegraphics[width=4.5in]{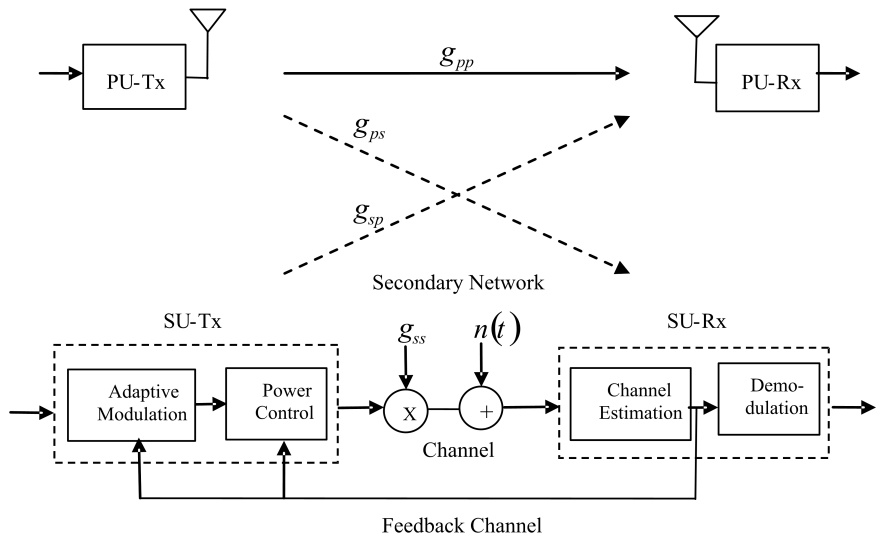}
\caption{System model of a Spectrum Sharing CRN}
\label{fig_SS_CRN}
\end{figure}

The secondary link employs adaptive modulation in conjunction with power control as depicted in Fig.2. A set of M-ary Quadrature Amplitude Modulations (M-QAM) is considered. The selection of a particular modulation and the optimal power allocation rely on the estimated SNR $\gamma_{ss}$ at the SU-Rx. As in the opportunistic spectrum access system model the symbol period is $T_s=1/B$ using ideal raised cosine pulses and the channel is a block flat fading channel model with a corresponding average received SNR at the SU-Rx, $\overline\gamma_{ss}$ \cite{Big01}.  

\subsection{Sensing-based SS CRN}
The PUs in SS CRNs can be further protected by employing a spectrum sensing mechanism that can efficiently identify resource availability \cite{Lee11}. In this case a new CRN model is defined known as sensing-based SS CRN \cite{Kang09}. We assume such a deployment in Fig.2 where a spectrum sensor at the SU-Tx determines the active or idle state of the PU.  In particular, the spectrum sensor acquires $N$ samples over an AWGN channel of the transmit signal at the PU-Tx, which is equal to $N=\tau f_s$  where $\tau$  is the sensing time and $f_s$ is the sampling frequency. By this procedure, the spectrum sensor senses a received SNR $ S $ produced by the PU-Tx that in sequel is compared with a sensing threshold $ n $ above which the PU is considered active.

If the PU is actually active and the sensing result is that the PU is active too, this scenario is known as perfect detection. The corresponding probability is referred to as probability of detection, which is denoted by $ d $. On the other hand, if the PU is actually inactive and the sensing result is active, this scenario is known as false alarm, and the corresponding probability is referred to as probability of false alarm, which is denoted by $ f $. In our model, the spectrum sensor is an energy detector which results in the following probabilities over an AWGN channel with zero mean and variance $ \sigma_n^2 $ \cite{Yue10}.
   
\begin{equation}
d=Q\left( \left( \frac{n}{\sigma_n^2}-S-1\right)\sqrt{\frac{\tau f_s}{2\gamma+1}} \right) 
\end{equation}

\begin{equation}
f=Q\left( \left( \frac{n}{\sigma_n^2}-1\right)\sqrt{\tau f_s} \right)
\end{equation} 

where $Q\left(\cdot\right)$ is the complementary distribution function of the standard Gaussian \footnote[1] {$Q\left(x\right)=1/\sqrt{2\pi}\int_{x}^\infty e^{\left( -t^2/2\right)}dt$}.
Therefore, the secondary link employs CR and DR adaptive modulation with MQAM constellations and power control in conjunction with spectrum sensing leading to a sensing-based SS CRN model. 
\section{Adaptive modulation in Cognitive Radio Networks over fading channels}
In this section, we provide the performance analysis regarding spectral efficiency and optimal power allocation when the CR and DR adaptive modulation schemes are implemented in OSA, SS and sensing-based SS CNRs.  
\subsection{Adaptive modulation in Opportunistic Spectrum Access CRNs}
\subsubsection{Continuous Rate adaptive modulation in OSA CRN}
The CR adaptive modulation with an MQAM constellation set results in the following expression for a specific bit error rate (BER) target value \cite{Gold97}
\begin{equation}
M(\gamma)=1+\frac{1.5\gamma}{-ln(5BER)}\frac{P(\gamma)}{\overline P} 
\end{equation}
where $ \gamma $ is the received SNR. On the other hand, the scope in OSA CRNs is to allocate a channel $ c $ to a user $ u $ which maximizes the average spectral efficiency (ASE). ASE is an expectation $E(\cdot)$ on the achievable symbol rate $log_2(M)$ \footnote[2]{The term average spectral efficiency is used when fading channels are assumed.} that is given as follows 
\begin{equation}
S_e=E[log_2M(\gamma)]=\int log_2\left( 1+\frac{1.5\gamma}{-ln(5BER)}\frac{P(\gamma)}{\overline P}\right) p(\gamma) d\gamma
\end{equation}
subject to the following power constraint 
\begin{equation}
\int P(\gamma)p(\gamma) d \gamma \leq \overline{P}.
\end{equation}
where $ p(\gamma) $ is the probability density function (PDF) of the received SNR $ \gamma $ \cite{Had08}. 
The optimal power allocation which maximizes the ASE in OSA CRNs with CR adaptive modulation is given from \cite{Gold97}
\begin{equation}
\frac{P(\gamma)}{\overline{P} }=\begin{cases} \frac{1}{\gamma_0}-\frac{1}{\gamma K} ,  & \gamma\geq \frac{\gamma_0}{K} \\
0, & \gamma < \frac{\gamma_0}{K}
\end{cases}
\end{equation}
where $ \gamma_0 $ is a cutoff value in the received SNR and $ K $ is the effective power loss that retains the BER value and it is equal to 
\begin{equation}
K=\frac{-1.5}{ln(5BER)}.
\end{equation}
Substituting (6) into (4), the ASE for the CR MQAM is maximized up to a cut-off level in SNR denoted as $ \gamma_K=\gamma_0/K $  that is derived from 
\begin{equation}
\langle Se \rangle_{CR}=\int_{\gamma_K}^\infty log_2\left( \frac{\gamma }{\gamma_K }\right) p(\gamma) d \gamma.
\end{equation}
In the considered OSA CRN, equation (8) gives the ASE that the PU achieves at a channel $ c $ denoted as $ Se_{1,c} $  since it is served first from the OSA strategy \cite{Had08}. Thus, the achieved ASE by a SU $ u $  denoted as $ Se_{u,c} $  is equal to 
\begin{equation}
Se_{u,c}=\Delta_{1,c} Se_{1,c}
\end{equation}
where $ \Delta_{1,c} $ is the spectrum factor gain which represents the probability that the channel $ c $ is not occupied by the PU. This gain depends on the cut-off level in SNR $ \gamma_k $ of the optimal power allocation over the fading channel and is given by 
\begin{equation}
\Delta_{1,c}=\int_0^{\gamma_K} p(\gamma) d \gamma .
\end{equation}
If we generalize this strategy for $ U $ users, the sum ASE which an OSA CRN provide is given as follows  
\begin{equation}
Se_{sum}=\sum_{u=1}^U Se_{u,c}={\sum_{u=1}^U \Delta_{1,c} Se_{1,c}}= \frac{1-\Delta_{1,c}^U }{1-\Delta_{1,c}} Se_{1,c}.
\end{equation}
The term $ {1-{\Delta^{U}_{1,c}}/1-{\Delta_{1,c}}} $  \footnote[3]{where $\sum_{u=1}^U x = 1-x^U/1-x$}  is called total band factor gain and it represents the percentage of the channels that remain unused and  can be exploited by the SUs which are served with a specific priority by the OSA CRN.  
\subsubsection{Discrete Rate adaptive modulation in OSA CRN}
We now consider a discrete rate (DR) MQAM with a constellation set of size $ N $ with  $ M_0=0, M_1=2 $ and  $ M_j=2^{2\left( j-1\right)} $   for $ j=2,...,N $. At each symbol time, the system transmits using a constellation from the set  $ \left\lbrace M_j=0,1,...,N\right\rbrace  $ \cite{Gold97}. The choice of a constellation depends on the fade level $ \gamma $  i.e. the SNR over that symbol time while the $ M_0 $ constellation corresponds to no data transmission. Therefore, in OSA CRNs, for each value of  $ \gamma $, the SU-Tx decides which constellation $ M $ to employ and what is the associated transmit power $ P $  in order to maximize the ASE. The ASE is in this case defined as the sum of the data rates of each constellation $j$ multiplied with the probability that this constellation will be selected i.e. $( \gamma_j \leq \gamma \leq \gamma_{j+1})$ and it is given from
\begin{equation}
\left\langle S_e \right\rangle _{DR}=\sum_{j=1}^N{log_2 ( M_j) p ( \gamma_j \leq \gamma \leq \gamma_{j+1}) }
\end{equation}
subject to the following power constraint
\begin{equation}
\sum_{j=1}^N \int_{\gamma_j}^{\gamma_{j+1}} \frac{P_j(\gamma)}{\overline{P}}p(\gamma) d\gamma=1
\end{equation}
where ${P_j(\gamma)}/{\overline{P}}$ is the optimal power allocation that is obtained from (3) for each constellation  $ M_j $ with a fixed BER. Thus the optimal power allocation is given from
\begin{equation}
\frac{P_j(\gamma)}{\overline{P} }=\begin{cases} (M_j-1) \frac{1}{\gamma K},  & M_j\leq \frac{\gamma}{\gamma^*} \leq M_{j+1} \\
0, & M_j=0
\end{cases}
\end{equation}
where $ \gamma^* $  is the cut-off level in SNR of the optimal power allocation which optimizes the amount of the fading regions $ \gamma_j $  for $ j=0,1,...,N $  according to $ \gamma_j=\gamma^* M_j $.  In this way the maximization in spectral efficiency is accomplished. Therefore, the band factor gain and the sum ASE in OSA CRNs which implement a DR adaptive modulation depend on the cut-off level in SNR,  $ \gamma^* $  and they are obtained by equations (10) and (11) accordingly. 

\subsection{Adaptive modulation in Spectrum Sharing CRNs}
\subsubsection{Continuous rate adaptive modulation in SS CRNs}
In SS CRNs, the implementation of the adaptive modulation aims to maximize the ASE while the PU is protected by an optimal power allocation policy \cite{Mus09}. Thus the SU-Tx employs power control that is based on an average transmit power constraint $ \overline{P} $ and a peak interference power constraint,  $ I_{pk} $ \footnote[4]
{Although, different optimal power allocation strategies for the SU can be found in [6], we assume this pair of constraints since it is the most representative in CRNs.} for the SU-Rx and for the PU-Rx respectively.
 
The transmit power is now related to both the received SNR at the SU-Rx,  $ \gamma_{ss} $ and the received SNR at the PU-Rx,  $ \gamma_{sp} $ which is denoted as $ P(\gamma_{ss},\gamma_{sp}) $. We consider also that the interference from PU-Tx to SU-Rx with channel power gain $ g_{ps} $  is negligible. Thus, when CR adaptive modulation is implemented in the SS CRN, the maximization of the ASE at the secondary link is obtained as follows   
\begin{equation}
\left\langle S_e \right\rangle _{CR}=\int_{0}^{\infty} \int_{0}^{\infty} log_2  \left( 1+\frac{1.5 \gamma_{ss}}{-ln(5BER)} \frac{P(\gamma_{ss}, \gamma_{sp})}{\overline{P} }\right) p(\gamma_{ss})p(\gamma_{sp}) d \gamma_{sp} d \gamma_{ss}
\end{equation}
subject to the following power constraints
\begin{equation}
\int_{0}^{\infty} \int_{0}^{\infty} P(\gamma_{ss}, \gamma_{sp}) p(\gamma_{ss})p(\gamma_{sp}) d \gamma_{sp} d \gamma_{ss} \leq \overline{P}
\end{equation}
\begin{equation}
\gamma_{sp} P(\gamma_{ss}, \gamma_{sp})\leq I_{pk}.
\end{equation}
where $ p(\gamma_{ss}) $ and $ p(\gamma_{sp}) $ are the PDFs of the corresponding received SNR. 
The optimal power allocation that maximizes equation (15) subject to (16) and (17) is given from 
\begin{equation}
\frac{P(\gamma_{ss}, \gamma_{sp})}{\overline{P} }= \begin{cases} \frac{1}{\gamma_{ss}^*}-\frac{1}{\gamma_{ss} K} ,  & \gamma_{ss}> \frac{\gamma_{ss}^*}{K} \;\; and \;\;   \gamma_{sp}< \frac{I_{pk}}{(1/{\gamma_{ss}^*}- 1/{\gamma_{ss} K})}  \\
\frac{I_{pk}}{\gamma_{sp}} , & \gamma_{ss}>\frac{\gamma_{ss}^*}{K} \;\; and \;\;   \gamma_{sp}\geq \frac{I_{pk}}{(1/{\gamma_{ss}^*}- 1/{\gamma_{ss} K})}
\end{cases}
\end{equation}
where $ {\gamma_{ss}^*}/{K} $ is the cut-off level in SNR above which the SU transmits in the shared band. On the other hand for $ \gamma_{ss}<{\gamma_{ss}^*}/{K} $  no data transmission occurs i.e. $ {P(\gamma_{ss}, \gamma_{sp})}=0 $ \cite{Xin09}.   
\subsubsection{Discrete rate adaptive modulation in SS CRNs}
In the considered SS CRN we assume now a DR adaptive modulation in conjunction with an interference power constraint for the PU's protection and an average power constraint at the SU as in the previous subsection. Then the maximization of the ASE is obtained as follows 
\begin{equation}
\left\langle S_e \right\rangle _{DR}=\sum_{j=1}^N log_2(M_j) \int_{\gamma_{ss,j}}^{\gamma_{ss,j+1}} \int_{0}^{\infty} p(\gamma_{ss})p(\gamma_{sp}) d \gamma_{sp} d \gamma_{ss}
\end{equation}
subject to the following constraints
\begin{equation}
\sum_{j=1}^N  \int_{\gamma_{ss,j}}^{\gamma_{ss,j+1}} \int_{0}^{\infty} P_j (\gamma_{ss}, \gamma_{sp}) p(\gamma_{ss})p(\gamma_{sp}) d \gamma_{sp} d \gamma_{ss} \leq \overline{P}
\end{equation}
\begin{equation}
\gamma_{sp} P_j (\gamma_{ss}, \gamma_{sp}) \leq I_{pk}.
\end{equation}
The optimal power allocation that maximizes (19) subject to (20) and (21) is given in this case as follows 
\begin{equation}
\frac{P_j(\gamma_{ss}, \gamma_{sp})}{\overline{P} }= \begin{cases}  \frac{(M_j-1)}{\gamma_{ss,j}{K}} ,  & \gamma_{ss}^* M_j \leq \gamma_{ss} \leq \gamma_{ss}^*{M_{j+1}} \;\; and \;\;   \gamma_{sp}< \frac{I_{pk}}{(M_j- 1/{\gamma_{ss,j} K})}  \\
\frac{I_{pk}}{\gamma_{sp}} , & \gamma_{ss}^* M_j \leq \gamma_{ss} \leq \gamma_{ss}^*{M_{j+1}} \;\; and \;\;   \gamma_{sp}\geq \frac{I_{pk}}{(M_j- 1/{\gamma_{ss,j} K})}
\end{cases}
\end{equation}
where $ \gamma_{ss}^* $  is the cut-off level in SNR of the optimal power allocation which designates the fading regions $ \gamma_{ss,j} $  for $ j=0,1,...,N $ according to  $ \gamma_{ss,j}= \gamma_{ss,j} M_j $. For $  M_j=0 $ data transmission does not take place i.e. ${P_j(\gamma_{ss}, \gamma_{sp})}=0$.

\subsection{Adaptive modulation in sensing-based SS CRNs}
In the sensing-based SS CRN, the SU-Tx adapts its transmit power in accordance to the inferred PU's state. Specifically, if the PU is considered active, the SU-Tx transmits with power $ P^1 $ and if it is considered idle the SU-Tx will transmit with power $ P^0 $. In order to care for the PU's protection it is is assumed that $ P^0>P^1 $ \cite{Kang09}. Thus, when adaptive modulation is implemented in a sensing-based SS CRN, the achievable spectral efficiency is separated into $ Se^0 $  and  $ Se^1 $ and it holds that $ Se^0>Se^1 $. If $ \pi_0 $ is the probability that the PU is idle and $ \pi_1 $ the probability that the PU is active, then the ASE that includes all four possible scenarios of spectrum sensing (detection, missed detection, false alarm and no false alarm) denoted as $ d, (1-d), f $  and   $ 1-f $ respectively is given as follows
\begin{equation}
\langle Se \rangle_{sens}=\pi_0 (1-f)Se^0 + \pi_0 f Se^1+ \pi_1(1-d) Se^0+\pi_1 d Se^1.
\end{equation}
In the next subsections we will study the problem of optimal power allocation for both CR and DR adaptive modulations which determine the transmit powers $ P^0 $  and  $ P^1 $ and in consequence the average spectral efficiencies $ Se^0 $  and  $ Se^1 $.
\subsubsection{Continuous rate adaptive modulation in sensing-based SS CRNs}
The ASE in CR adaptive modulation in SS CRNs is provided by (15). Thus, assuming transmit powers $ P^0(\gamma_{ss}, \gamma_{sp}) $, $ Se^0 $  and  $ Se^1 $ are given by
\begin{equation}
P^0(\gamma_{ss}, \gamma_{sp})=\frac{1}{\gamma_{ss}^*}-\frac{1}{\gamma_{ss}K}\;\;for\;\;\gamma_{ss}>\gamma_{ss}^*/K
\end{equation}
where represents the well known optimal power allocation based on water-filling algorithm and $ P^1(\gamma_{ss}, \gamma_{sp}) $ given by
\begin{equation}
\frac{P^1(\gamma_{ss}, \gamma_{sp})}{\overline{P} }= \begin{cases} \frac{1}{\gamma_{ss}^*}-\frac{1}{\gamma_{ss} K} ,  & \gamma_{ss}> \frac{\gamma_{ss}^*}{K} \;\; and \;\;   \gamma_{sp}< \frac{I_{pk}}{(1/{\gamma_{ss}^*}- 1/{\gamma_{ss} K})}  \\
\frac{I_{pk}}{\gamma_{sp}} , & \gamma_{ss}>\frac{\gamma_{ss}^*}{K} \;\; and \;\;   \gamma_{sp}\geq \frac{I_{pk}}{(1/{\gamma_{ss}^*}- 1/{\gamma_{ss} K})}
\end{cases}
\end{equation}
where $ {\gamma_{ss}^*}/{K} $  is the cut-off level in SNR above which the SU transmits through the shared band and $ {\gamma_{sp}} $ is the SNR level at the interference channel above which the transmit power is truncated by the interference power constraint $I_{pk}$. In both (24) and (25) for $ \gamma_{ss}<{\gamma_{ss}^*}/{K} $  no data transmission occurs and thus $ {P^0(\gamma_{ss}, \gamma_{sp})}={P^1(\gamma_{ss}, \gamma_{sp})}=0 $.  
\subsubsection{Discrete rate adaptive modulation in sensing-based SS CRNs}
Relying on (19), the average spectral efficiencies $ Se^0 $  and  $ Se^1 $ for DR adaptive modulation in sensing-based SS CRNs assuming transmit powers $ P_j^0(\gamma_{ss}, \gamma_{sp}) $  and  $ P_j^1(\gamma_{ss}, \gamma_{sp}) $ for each constellation $ M_j $  when the PU is idle or active respectively are given by
\begin{equation}
P_j^0(\gamma_{ss}, \gamma_{sp})=(M_j-1) \frac{1}{\gamma_{ss,j}K},\;\;for\;\;\gamma_{ss}^*M_j \leq \gamma_{ss} \leq \gamma_{ss}^*M_{j+1}
\end{equation}
\begin{equation}
\frac{P_j^1(\gamma_{ss}, \gamma_{sp})}{\overline{P} }= \begin{cases}  \frac{(M_j-1)}{\gamma_{ss,j}{K}} ,  & \gamma_{ss}^* M_j \leq \gamma_{ss} \leq \gamma_{ss}^*{M_{j+1}} \;\; and \;\;   \gamma_{sp}< \frac{I_{pk}}{(M_j- 1/{\gamma_{ss,j} K})}  \\
\frac{I_{pk}}{\gamma_{sp}} , & \gamma_{ss}^* M_j \leq \gamma_{ss} \leq \gamma_{ss}^*{M_{j+1}} \;\; and \;\;   \gamma_{sp}\geq \frac{I_{pk}}{(M_j- 1/{\gamma_{ss,j} K})}
\end{cases}
\end{equation}
where $ \gamma_{ss}^* $  is the cut-off level in SNR of the optimal power allocation which optimizes the fading regions $ \gamma_{ss,j} $  for $ j=0,1,...,N $ according to  $ \gamma_{ss,j}= \gamma_{ss}^* M_j $ .  In both (26) and (27) and for $  M_j=0 $  no data transmission occurs and thus ${P_j^0(\gamma_{ss}, \gamma_{sp})}={P_j^1(\gamma_{ss}, \gamma_{sp})}=0$.
We have to note here that the ASE in (23) is equal to $ (T-\tau)/T\langle Se \rangle_{sens} $ when a frame of  duration $ T $  is transmitted over the sensed and subsequently shared band which is known as throughput $\xi$ measured in bits per seconds per hertz. Thus, the aim in a sensing based SS CRN is to maximize the throughput  $\xi$ on the transmit powers  $ P^0 $  and  $ P^1 $  with a constraint on sensing time $ \tau $  or sensing threshold $ \eta $  \cite{Kang09} \cite{Fou10}. Although this problem is concave on these three parameters i.e. transmit powers, sensing time and threshold, due to the complicated coupling expression of ASE in (23) the solution is obtained through an iterative algorithm as presented in\cite{Kang09}. Thus, for both constant sensing time and threshold \footnote[5]{We assume constant sensing time and threshold since in this paper we do not consider the maximization on these two parameters.}, the iterative algorithm results in a common cut-off level
  $ \gamma_{ss}^* $  in the received SNR at the SU-Rx which provides the optimal power allocation in the secondary link according to the transmit power constraints and the PU's activity. In this way we obtain numerical results regarding the average spectral efficiencies $Se^0 $ and  $ Se^1 $ as well as the corresponding optimal power allocation as defined in equations (24) through (27) for CR and DR adaptive modulations in sensing-based SS CRNs. Details regarding this particular optimization problem and the corresponding iterative algorithm that is based on the subgradient method can be found in \cite{Kang09} that we omit here since it is out of our scope in this paper. Notably, we do not consider any changes on the probability of PU's activity $ \pi_0 $ within the time period of the frame's duration $ T $  but we assume this variable as constant. On the other hand, a practical example that considers a stochastic model for the PU's activity can be found in \cite{Ren09} which we aim to take into account in a future investigation on this topic.  
  
\section{Results and discussion}

In Fig.3 we present the results obtained when CR adaptive modulation in OSA CRN over fading channel is considered. We assume a Rayleigh distribution for the fading channel with PDF equal to $ {1 / \overline{\gamma}}exp(-\gamma / \overline{\gamma}) $  where  $ \gamma $ and $ \overline{\gamma} $  are the instantaneous and the average received SNR respectively. We depict the results for BER equal to  $ 10^{-3} $ and $ 10^{-6} $. With solid lines are shown the results for the case of PN only, which is the case of a conventional network that does not serve any SUs. With dashed lines are shown the results obtained for the OSA CRN with a number of users equal to $ U=5 $ i.e. one PU and four SUs. An important performance gain in the OSA CRN is observed in comparison with the performance of the conventional network i.e. with $ U=1 $. For both BER  cases  the additional ASE is close to $ 0.5bits/sec/Hz $   at low average SNR regions e.g. $ \overline{\gamma}=0dB $. Besides, the additional ASE is close to  $ 0.3bits/sec/Hz $ at moderate average SNR regions e.g. $ \overline{\gamma}=10dB $  and finally the ASE is close to  $ 0.1bits/sec/Hz $   at high average SNR regions e.g.  $ \overline{\gamma}=20dB $. This behavior in particular, is explained from the fact that the probability that a channel is not allocated to the PU is larger for low average SNRs. In other words, for the low average SNR regions, the cut-off level in SNR  $ \gamma_K $ is getting higher and in consequence the band factor gain in equation (10) is getting higher too. The opposite holds for the high average SNRs where the PU is more likely to transmit over a channel since the cut-off level in SNR $ \gamma_K $  is getting lower and thus the constraint for allocating a channel is relaxed.

\begin{figure}[tpb]
\centering
\includegraphics[width=5.5in, height=3.5in]{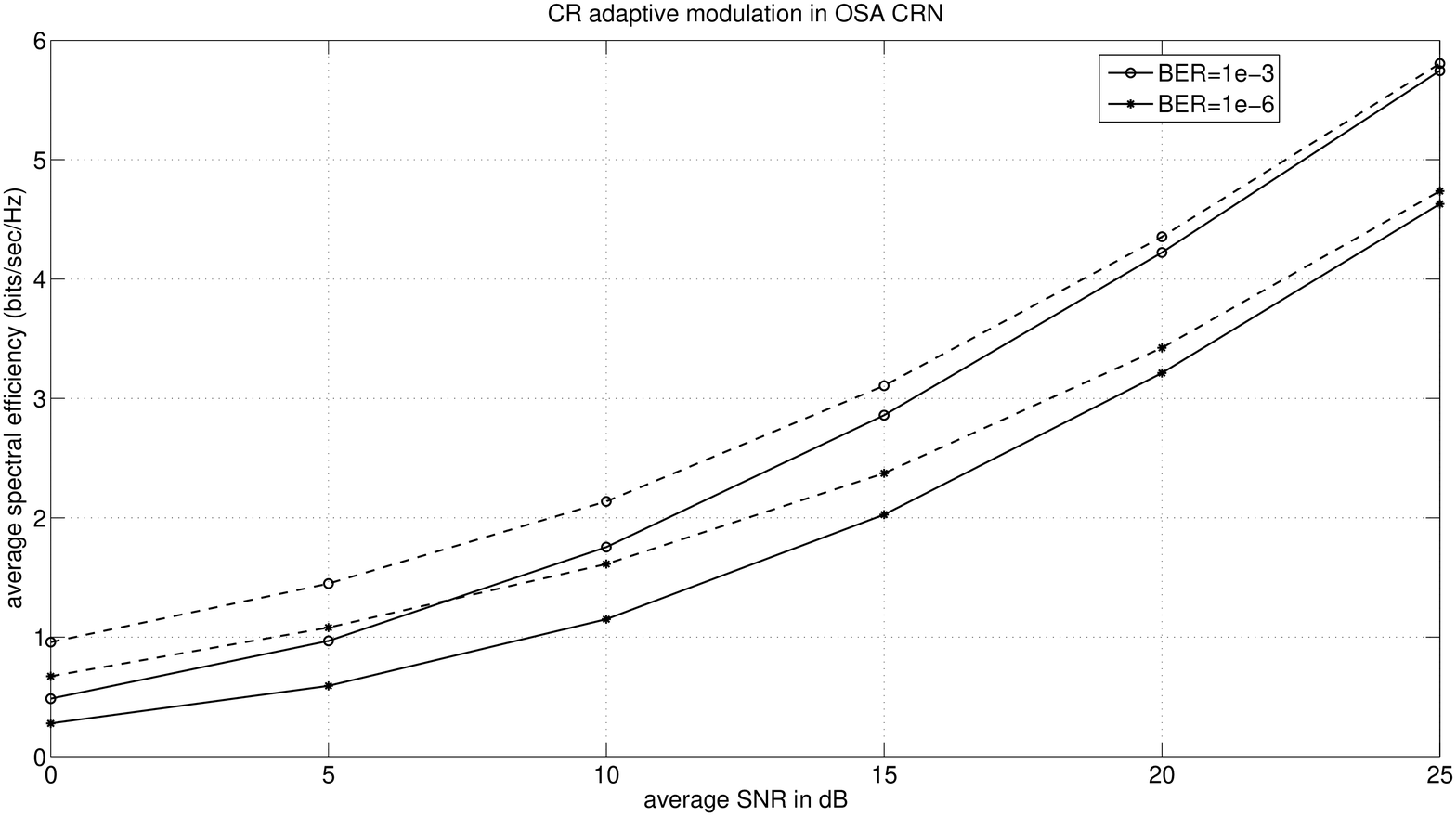}
\caption{Average spectral efficiency (ASE) of CR adaptive modulation in OSA CRN with $ U=5 $ users (dashed lines) over Rayleigh fading channel and $ U=1 $ users (solid lines)}
\label{fig_ASE_in_OSA}
\end{figure}

Fig.4a and Fig.4b show the results obtained when the discrete rate (DR) adaptive modulation in OSA CRN over a Rayleigh fading channel is considered. The average received SNR follows the aforementioned Rayleigh distribution and is denoted by $ \overline{\gamma} $. We depict the results for a BER equal to and $ 10^{-3} $  in Fig.4a and the results for a BER equal to $ 10^{-6} $  in Fig.4b. With solid lines are shown the results for the conventional network i.e. for a PU  $ U=1 $ and with dashed lines are shown the results obtained for the OSA CRN with a total number of users equal to  $ U=5 $ as before. We consider 5 fading regions with a set of MQAM constellations $ \left\lbrace 0,2,4,16,64\right\rbrace  $ , 4  fading regions with a set of MQAM constellations $ \left\lbrace 0,2,4,16\right\rbrace  $  and 3 fading regions with a set of MQAM constellations  $ \left\lbrace 0,2,4\right\rbrace  $. Again, the performance gain is remarkable for low average SNR regions and is due to the increase in the probability of finding an unoccupied channel as mentioned previously for the OSA CRN.  It should be noted that the performance gain is identical at low average SNR regions for all fading regions i.e. 5,4 and 3 and it is close to  $ 0.3bits/sec/Hz $. On the other hand, the performance gain is negligible at high average SNR regions due to the decreased possibility of finding a channel opportunity. Regarding the different BER values, the tighter the BER is get i.e. $ BER=10^{-6} $ , the larger the performance gain is become, something that we discuss in detail in Fig.5 which illustrates the total band factor gain for the CR and DR implementations of adaptive modulation in OSA CRNs. 

\newcounter{subfigure1}
\renewcommand{\thefigure}{\arabic{figure}\alph{subfigure1}}
\setcounter{subfigure1}{1}

\begin{figure}[tpb]
\centering
\includegraphics[width=5.5in, height=3.5in]{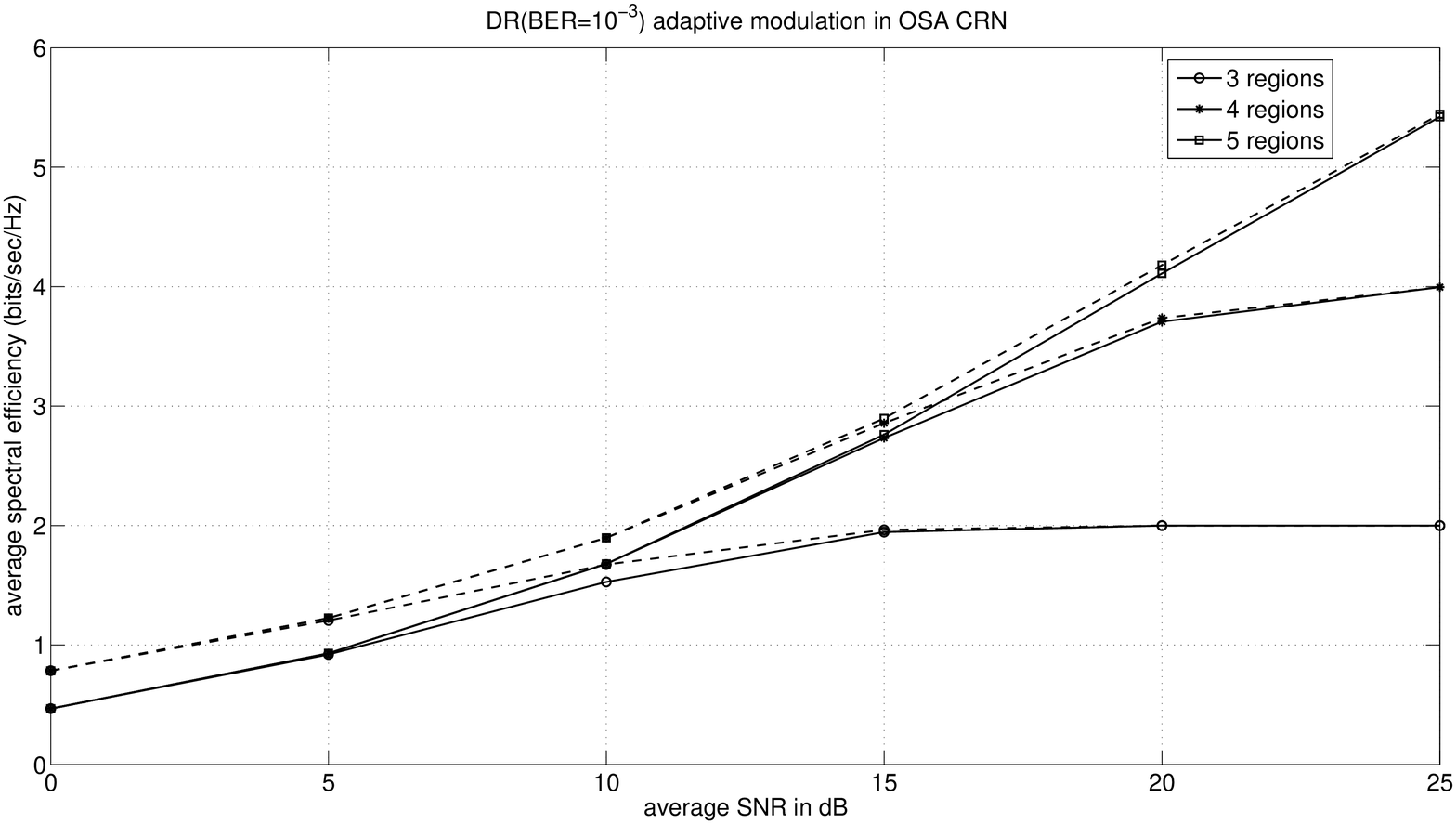}
\caption{Average spectral efficiency (ASE) of DR ($ BER=10^{-3}  $) adaptive modulation in OSA CRN with $ U=5 $ users (dashed lines) over Rayleigh fading channel and with $ U=1 $ users (solid lines)}
\label{fig_SS_CRN-3}
\end{figure}

\addtocounter{figure}{-1}
\addtocounter{subfigure1}{1}

\begin{figure}[tpb]
\centering
\includegraphics[width=5.5in, height=3.5in]{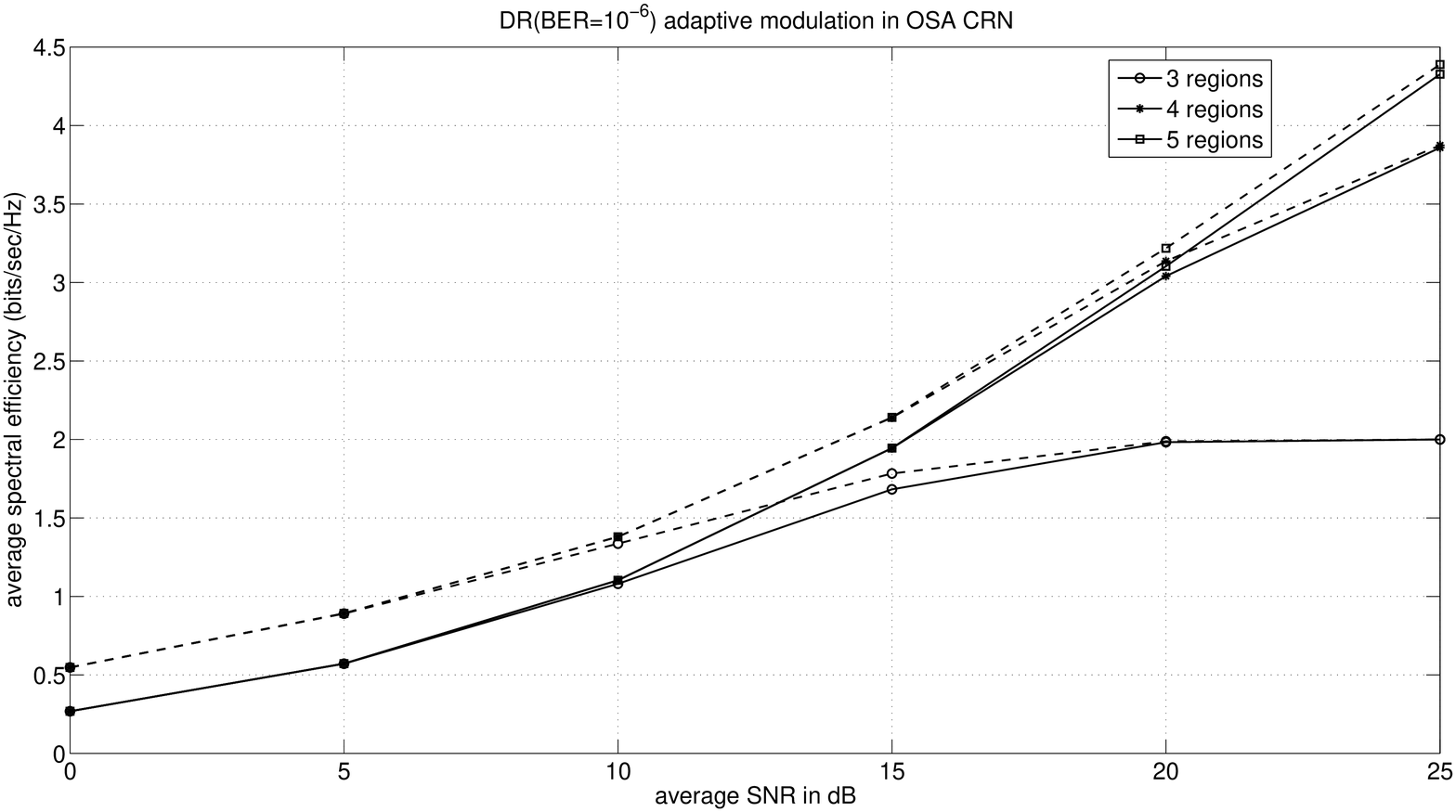}
\caption{Average spectral efficiency (ASE) of DR ($ BER=10^{-6}  $) adaptive modulation in OSA CRN with $ U=5 $ users (dashed lines) over Rayleigh fading channel and with $ U=1 $ users (solid lines)}
\label{fig_SS_CRN-6}
\end{figure}

\renewcommand{\thefigure}{\arabic{figure}}

Fig.5 shows the total band factor gain of adaptive modulation in OSA CRN as it is obtained from equation (11). We depict both CR and DR adaptive modulation cases over a Rayleigh fading channel. With solid lines are shown the results obtained for a BER equal to  $ 10^{-3} $ and with dashed lines are shown the results obtained for a BER equal to $ 10^{-6} $. The OSA CRN is considered with a number of users equal to  $ U=5 $. Notably, the largest total band factor gain is achieved in case of CR adaptive modulation with a BER equal to  $ 10^{-6} $ and the smallest one is achieved in case of DR adaptive modulation with 3 regions and a BER equal to $ 10^{-3} $ . Therefore, the tighter the BER criterion is become, the larger the advantage of the application of the OSA strategy in CRNs. We should further notice that the gain for  CR and DR adaptive modulations with a high number of regions i.e. 5 was expected due to the transmission with high bit rates in terms of bits per symbol and in consequence with a high ASE.  

\begin{figure}[tpb]
\centering
\includegraphics[width=5.5in, height=3.5in]{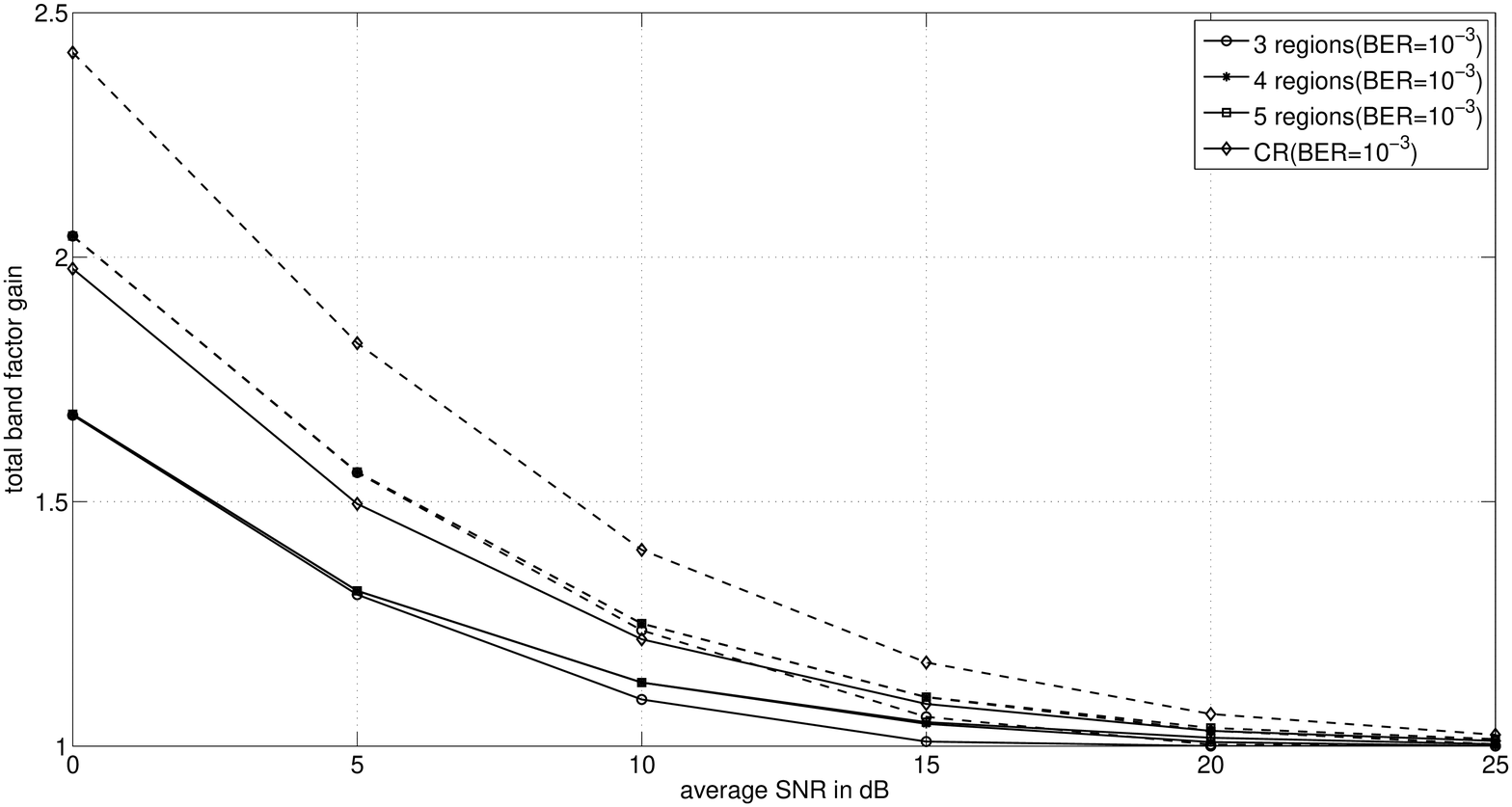}
\caption{Total band factor gain of DR adaptive modulation in OSA CRN  considering different number of fading regions with $ BER=10^{-3} $ (solid lines) and $ BER=10^{-6} $ (dashed lines) }
\label{Total_BFG}
\end{figure}

Fig.6 shows the results obtained when the CR adaptive modulation in the SS CRN over a Rayleigh fading channel is considered. We assume again the aforementioned Rayleigh distribution with $ \overline{\gamma} $  as the average received SNR. The results are obtained for different peak interference constraint values. We depict also the case without a peak interference constraint that corresponds to the conventional power control scheme \cite{Al99}. With solid lines are shown the results obtained for a BER equal to  $ 10^{-3} $ and with dashed lines are shown the results obtained for a BER equal to  $ 10^{-6} $. The gain is obvious when the BER value is less demanding i.e. equal to $ 10^{-3} $. However, there is no gain from adaptive modulation for an average SNR up to the interference power constraint i.e. $ \overline{\gamma}>I_{pk} $. This was expected since the transmit power $P(\gamma_{ss}, \gamma_{sp})$  is truncated by the power control in order to provide protection at the primary link and especially at the PU-Rx. Fig. 7 shows the results obtained when the discrete rate (DR) adaptive modulation in the SS CRN over a Rayleigh fading channel is considered. We assume the same channel distribution and peak interference constraint values with those considered for obtaining the results of fig. 6. With solid lines are depicted the results for 5 regions, with dashed lines are depicted the results for 4 regions and with dashed dotted lines we depict the results for 3 regions. Obviously, the performance gain of DR adaptive modulation is again truncated for protecting the PU and is decreasing when the BER target value is less demanding. 

\begin{figure}[tpb]
\centering
\includegraphics[width=5.5in, height=3.5in]{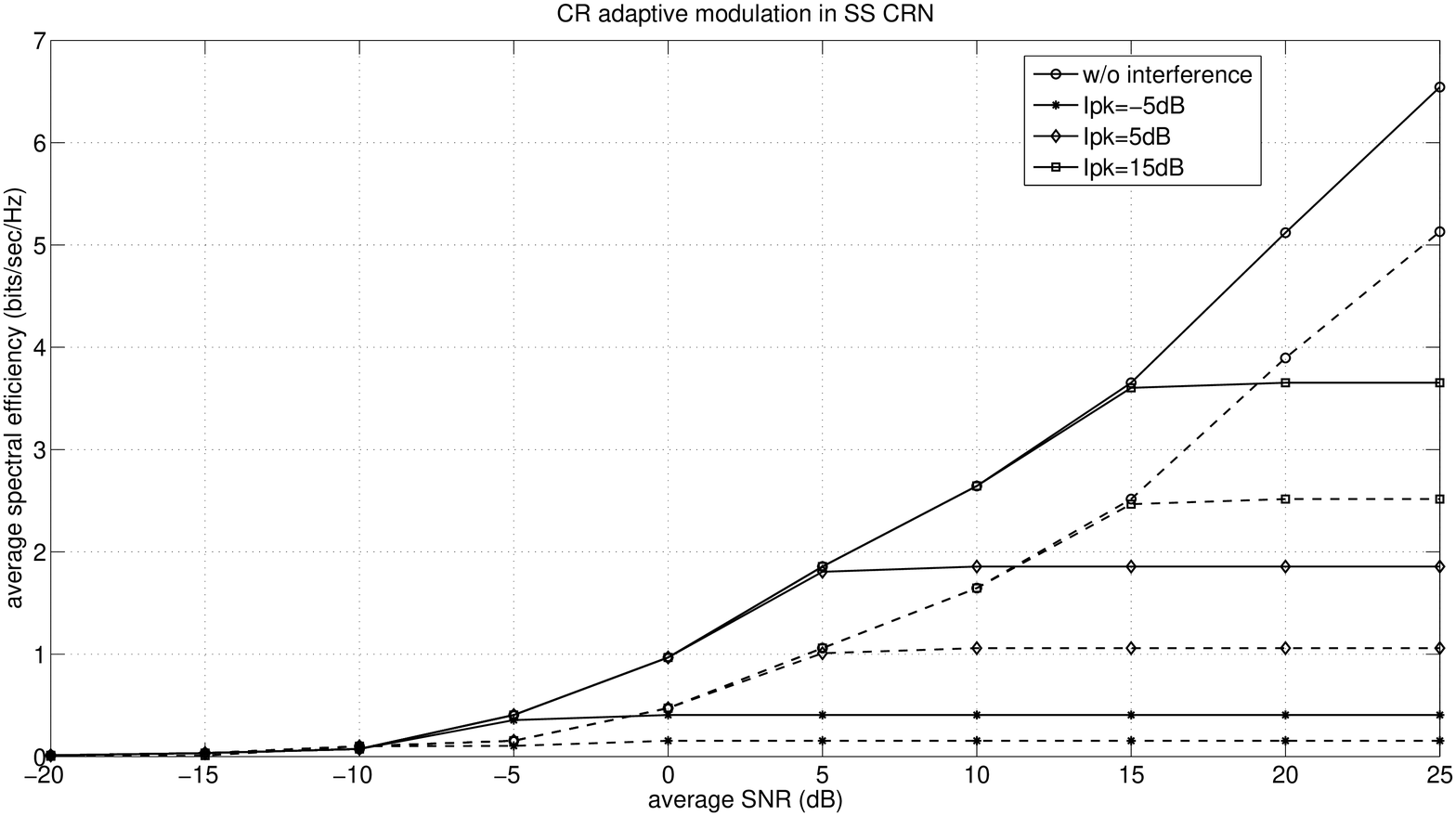}
\caption{Average spectral efficiency (ASE) of CR adaptive modulation in SS CRN over Rayleigh fading channel with different interference power constraints for $ BER=10^{-3} $ (solid lines) and $ BER=10^{-6} $ (dashed lines)}
\label{ASE_in_CR_SS_CRN}
\end{figure}

\begin{figure}[tpb]
\centering
\includegraphics[width=5.5in, height=3.5in]{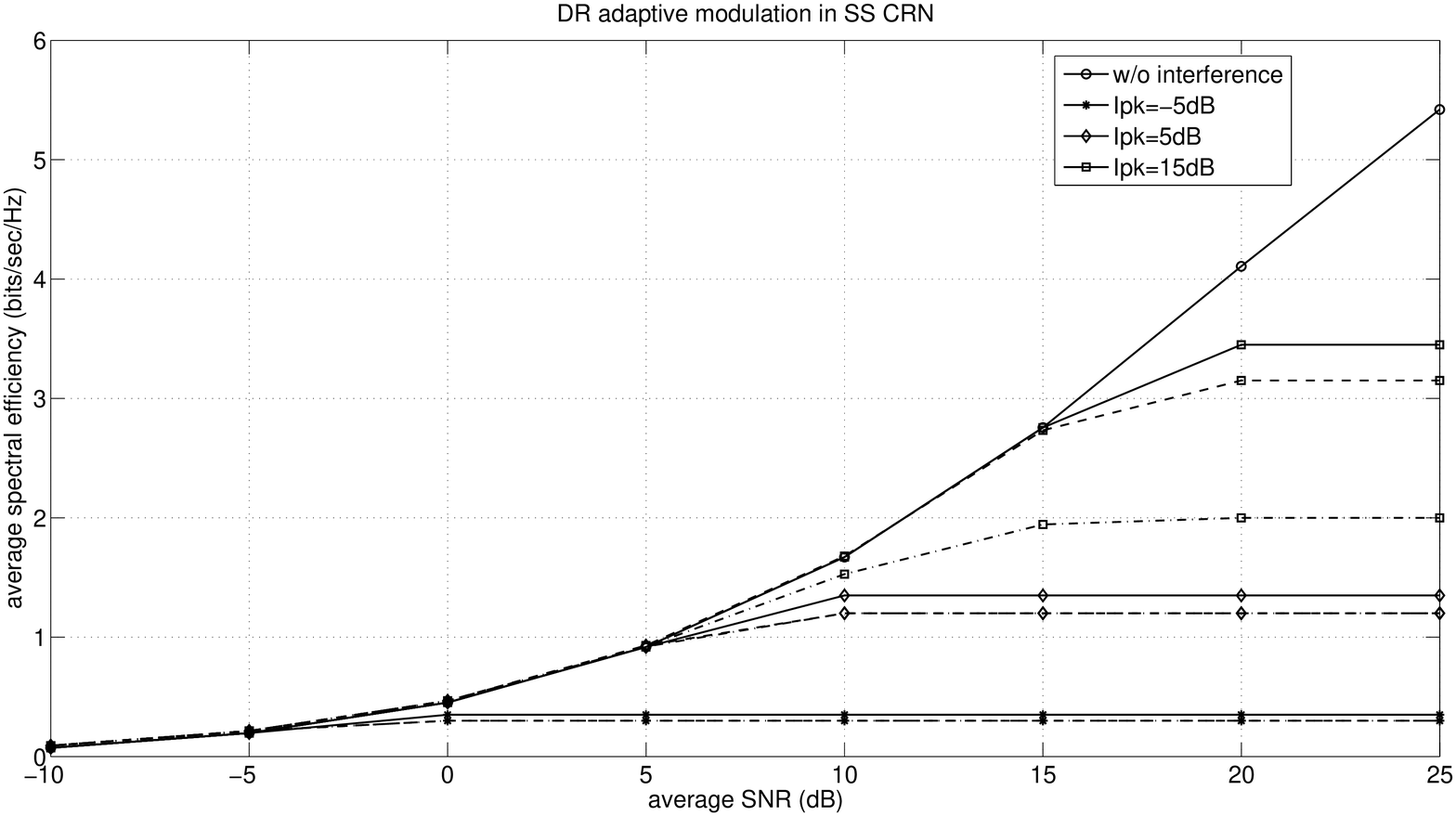}
\caption{Average spectral efficiency (ASE) of DR adaptive modulation in SS CRN over Rayleigh fading channel with different interference power constraints for  ($ BER=10^{-3} $) (solid lines) and ($ BER=10^{-6} $) (dashed lines) }
\label{ASE_in_DR_SS_CRN-3}
\end{figure}

Figures 8a, 8b and 8c show the results for the optimal power allocation which we derived for CR and DR adaptive modulations in OSA CRNs. Fig. 8a depicts the results for a Rayleigh fading channel with average SNR equal to  $ \overline{\gamma}=0dB $, fig. 8b depicts the case of an average SNR equal to  $ \overline{\gamma}=5dB $ and fig. 8c depicts the case of an average SNR equal to  $ \overline{\gamma}=15dB $. For all three cases, we suppose optimal power allocation of a CR adaptive modulation in an OSA CRN as obtained by equation (6) for $ BER=10^{-3} $ and $ BER=10^{-6} $. Besides, we suppose the optimal power allocation of a DR adaptive modulation in an OSA CRN as obtained by equation (14) for $ BER=10^{-3} $ and $ BER=10^{-6} $ considering 3, 4 and 5 regions respectively. It is obvious from all figures that the transmit power adaptation mechanism demands more power when a tight BER is required for both the CR and DR with 5 regions. On the opposite, the less consuming scenarios are the cases of a  $ BER=10^{-3} $   and DR with 3 regions. It should be noted that the power adaptation mechanism is not smooth in DR adaptive modulation and it presents an envelope with abrupt changes that reveal the transition to a new constellation size. Furthermore, comparing the power allocations for different average SNRs  $ \overline{\gamma} $, it is evident that the more the average SNR the less the transmit power consumption. It should be mentioned also that the optimal power allocation for the SS CRN is identical to OSA CRN when the average SNR is less than the peak interference power constraint i.e.  $ \overline{\gamma} \leq I_{pk} $ since in this case the conventional power control is applied as figured out from the results of ASE in SS CRNs described above. In the same context, for the cases that $ \overline{\gamma} \geq I_{pk} $ holds,  the optimal power allocation is identical to OSA CRN for $ \overline{\gamma}=I_{pk} $ since power is truncated at this level. Thus, in SS CRNs the optimal power allocation is less demanding when $ \overline{\gamma} \geq I_{pk} $ holds. Hence it does not permit any enhancement in terms of ASE as shown above which is sacrificed for protecting the PU’s transmission. 

\newcounter{subfigure2}
\renewcommand{\thefigure}{\arabic{figure}\alph{subfigure2}}
\setcounter{subfigure2}{1}

\begin{figure}[tpb]
\centering
\includegraphics[width=5.5in, height=3.5in]{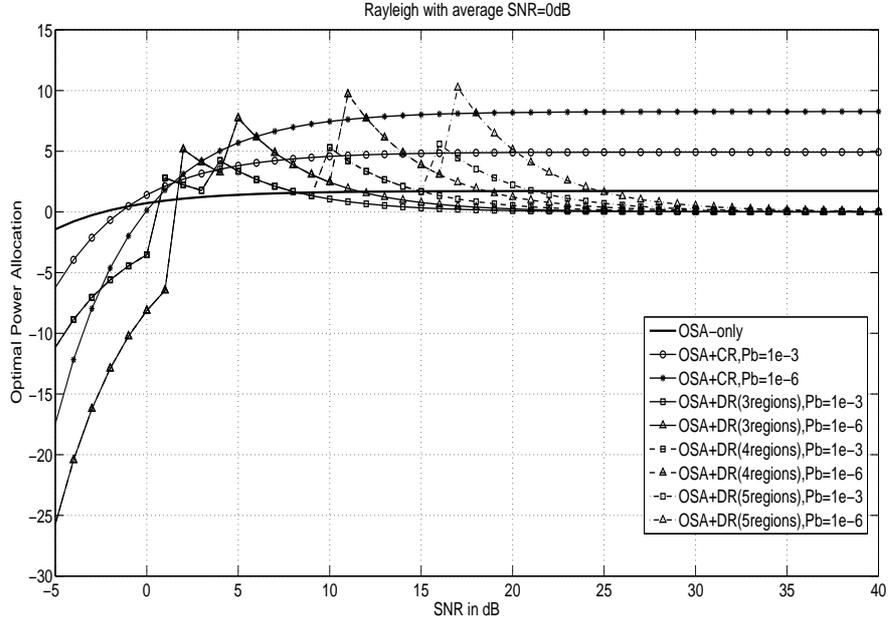}
\caption{Optimal power allocation $ P/ \overline{P} $  for an average SNR $ \overline{\gamma}=0dB $ in OSA CRNs for different CR and DR adaptive modulation schemes }
\label{OPA_0DB}
\end{figure}

\addtocounter{figure}{-1}
\addtocounter{subfigure2}{1}

\begin{figure}[tpb]
\centering
\includegraphics[width=5.5in, height=3.5in]{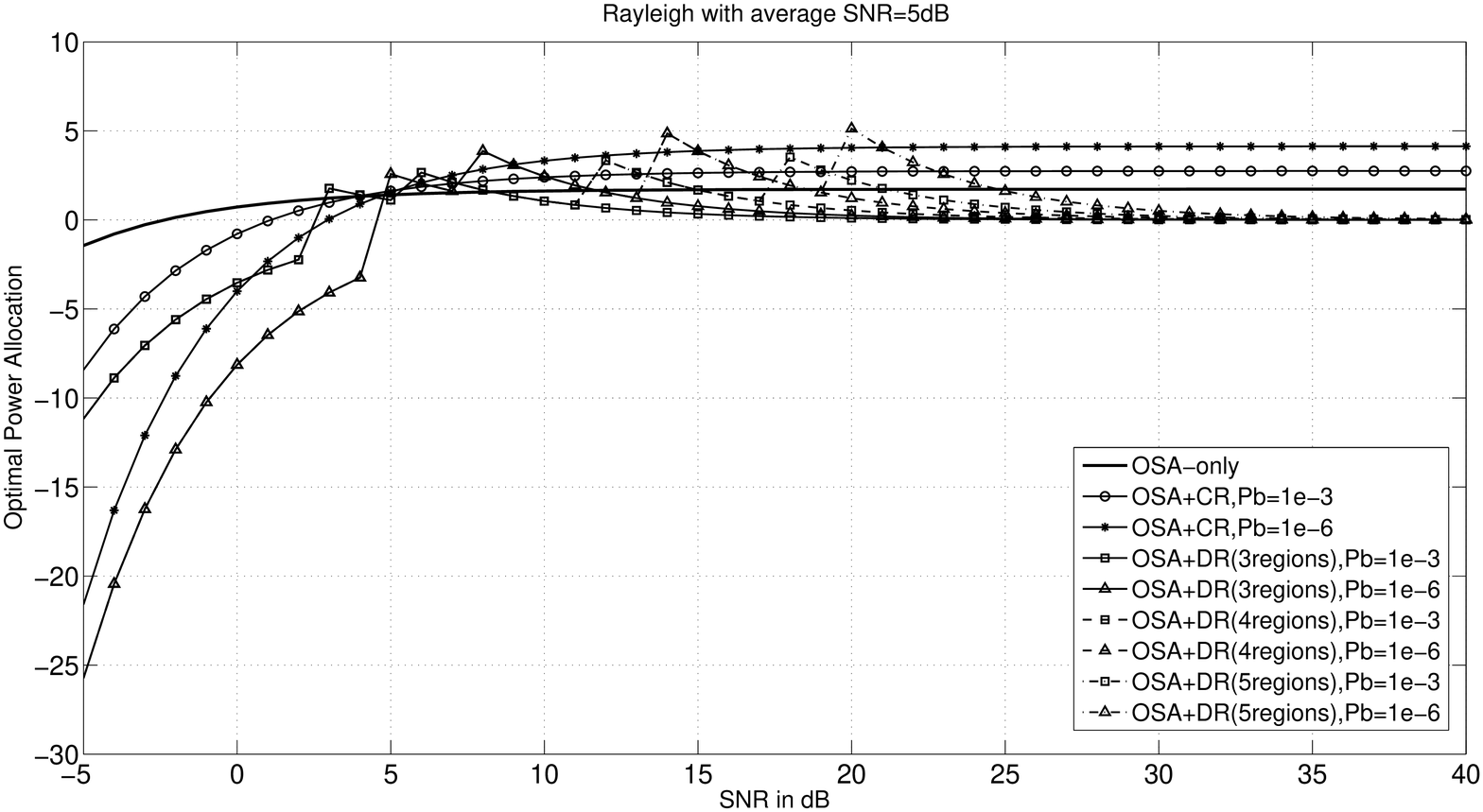}
\caption{Optimal power allocation $ P/ \overline{P} $  for an average SNR $ \overline{\gamma}=5dB $ in OSA CRNs for different CR and DR adaptive modulation schemes }
\label{OPA_5DB}
\end{figure}

\addtocounter{figure}{-1}
\addtocounter{subfigure2}{1}

\begin{figure}[tpb]
\centering
\includegraphics[width=5.5in, height=3.5in]{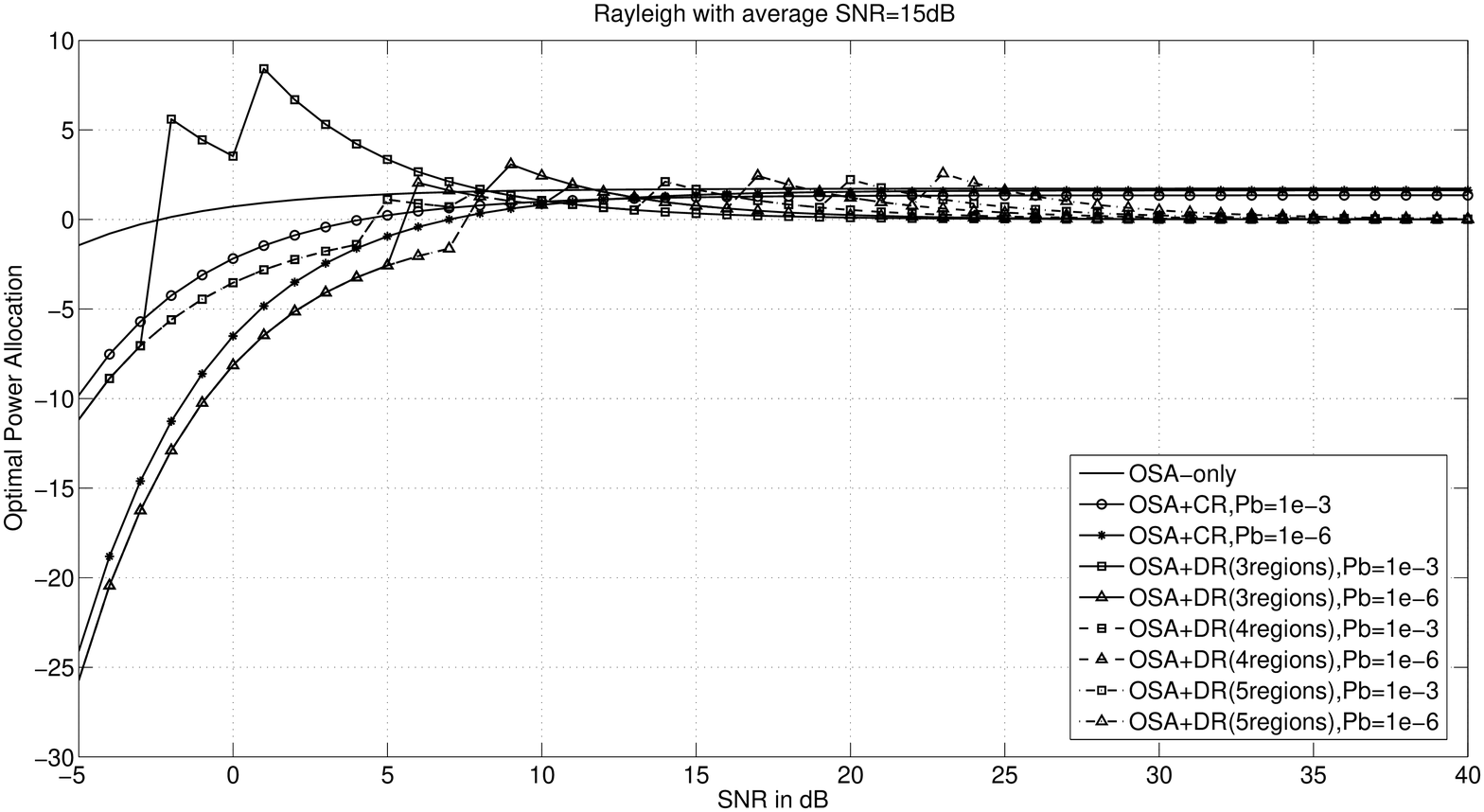}
\caption{Optimal power allocation $ P/ \overline{P} $  for an average SNR $ \overline{\gamma}=15dB $ in OSA CRNs for different CR and DR adaptive modulation schemes }
\label{OPA_15DB}
\end{figure}

\renewcommand{\thefigure}{\arabic{figure}}

Figures 9a and 9b show the results of the achievable throughput $ \xi_s $ at the SU-Tx for both CR and DR adaptive modulations schemes in sensing-based SS CRNs in order to highlight the impact of spectrum sensing mechanism on the performance of adaptive modulation. Both results are obtained for a BER equal to  $ 10^{-3} $. Spectrum sensing is considered for a frame duration equal to $ T=100ms $  and sensing time equal to  $ \tau=2ms $. PU's activity is determined from the probabilities  $ \pi_0=0.6 $ and $ \pi_1=0.4 $ which are considered constants within the period of frame duration $ T $. The probability of detection $ d $ is considered equal to $ 0.8 $, $ 0.5 $, $ 0.1 $ and $ 0.01 $, the sensed SNR  $ S $ is considered equal to $ -15dB $  and the standard deviation is taken equal to $ \sigma_n=1 $ . The peak interference power constraint $ I_{pk} $  is assumed equal to $ 0dB $  and   $ 10dB $. With dashed lines are depicted the results obtained when the peak interference power constraint is equal to  $ I_{pk}=0dB $  while the solid lines depict the results obtained when the peak interference power constraint is equal to $ I_{pk}=10dB $. The results show that the throughput is maximized when the probability of detection is increased and the interference is less demanding i.e. $ I_{pk}=10dB $. 

\newcounter{subfigure3}
\renewcommand{\thefigure}{\arabic{figure}\alph{subfigure3}}
\setcounter{subfigure3}{1}

\begin{figure}[tpb]
\centering
\includegraphics[width=5.5in, height=3.5in]{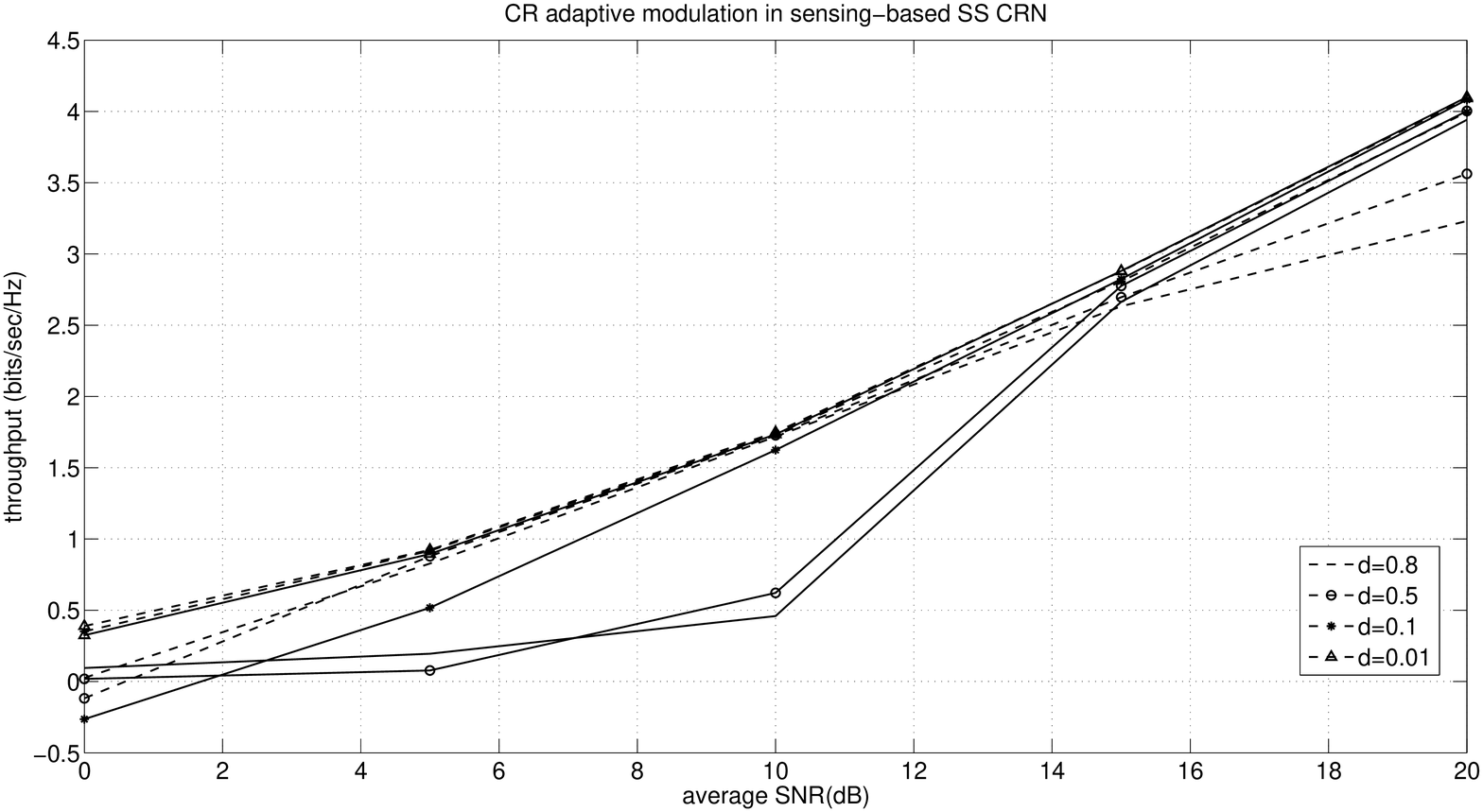}
\caption{Throughput $ \xi_s $ at the SU of CR adaptive modulation in sensing-based SS CRN over Rayleigh fading channel for different probabilities of detection $ d $  with peak interference power constraint $ I_{pk}=0dB $ (dashed lines) and $ I_{pk}=10dB $  (solid lines)}
\label{THR_CRAM}
\end{figure}

\addtocounter{figure}{-1}
\addtocounter{subfigure3}{1}

\begin{figure}[tpb]
\centering
\includegraphics[width=5.5in, height=3.5in]{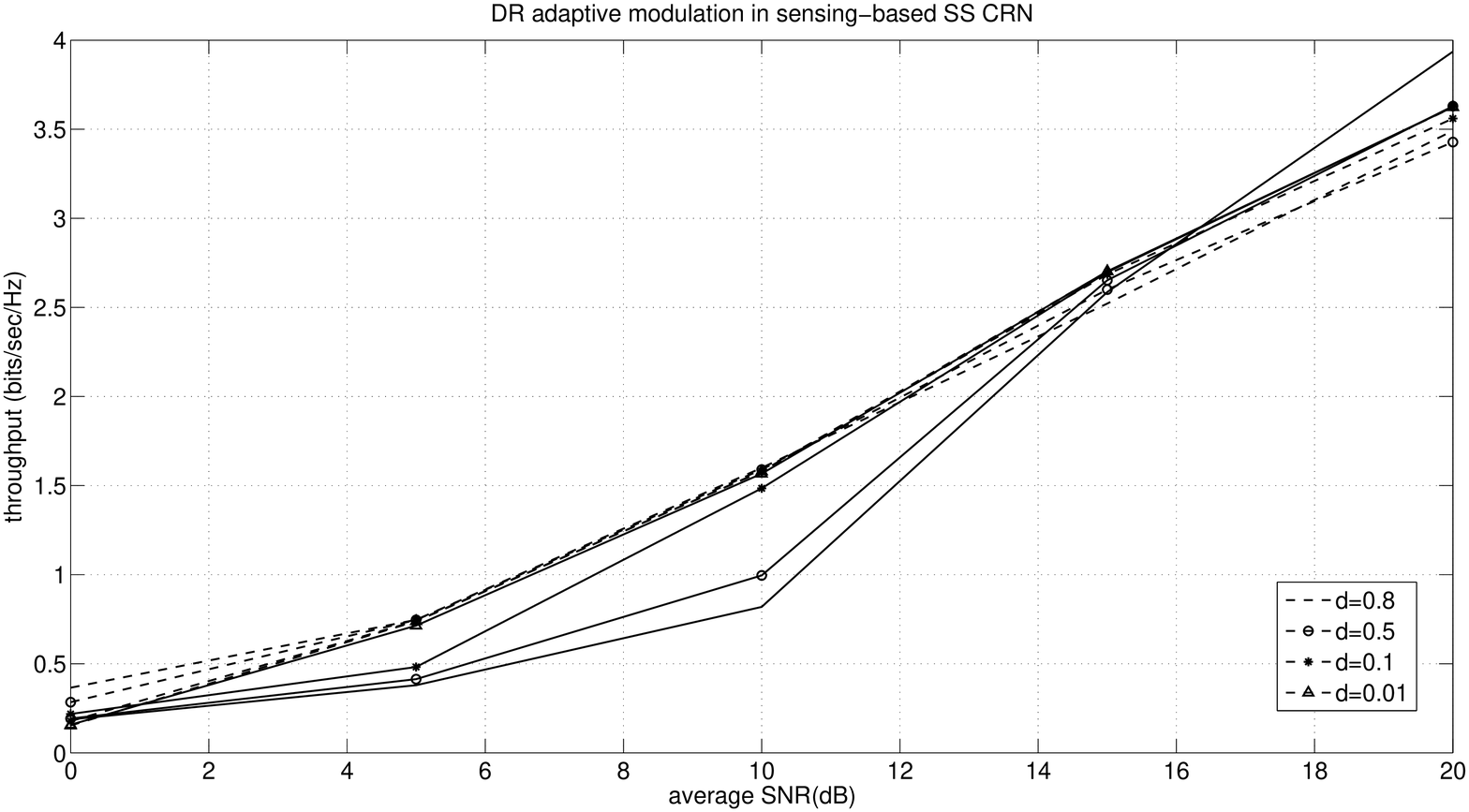}
\caption{Throughput $ \xi_s $ at the SU of DR adaptive modulation in sensing-based SS CRN over Rayleigh fading channel for different probabilities of detection $ d $  with peak interference power constraint $ I_{pk}=0dB $ (dashed lines) and $ I_{pk}=10dB $  (solid lines).}
\label{THR_DRAM}
\end{figure}

\renewcommand{\thefigure}{\arabic{figure}}

Specifically, the tighter the probability of detection  $ d $ e.g. $ d=0.8 $, the less the throughput maximization achieved. However, when the interference power constraint is getting tighter i.e.  $ I_{pk}=0dB $, the gain is maximized in average SNR regions. This is because a tighter interference power constraint reveals lower cut-off levels $ \gamma_{ss}^* $  in SNRs in relation to those achieved when a more loose constraint is considered. This is also evident from the optimal power allocation in fig.10a which shows that the power consumption is larger in the case of  $ I_{pk}=0dB $  than in the case of  $ I_{pk}=10dB $  due to the lower cut-off levels,  $ \gamma_{ss}^* $. Furthermore, a low value of probability of detection i.e.  $ d=0.01 $   requires more power consumption than a high one although it achieves better throughput maximization as it has been shown in fig.9a. An identical behavior is realized in the case of DR adaptive modulation in a sensing-based SS CRN as it is shown in figures 9b and 10b for the throughput and optimal power allocation respectively. The numerical results for this case are obtained for a constellation set of 5 regions i.e. $ \left\lbrace 0,2,4,16,64 \right\rbrace  $  with a BER equal to  $ 10^{-3} $.

\newcounter{subfigure4}
\renewcommand{\thefigure}{\arabic{figure}\alph{subfigure4}}
\setcounter{subfigure4}{1}

\begin{figure}[tpb]
\centering
\includegraphics[width=5.5in, height=3.5in]{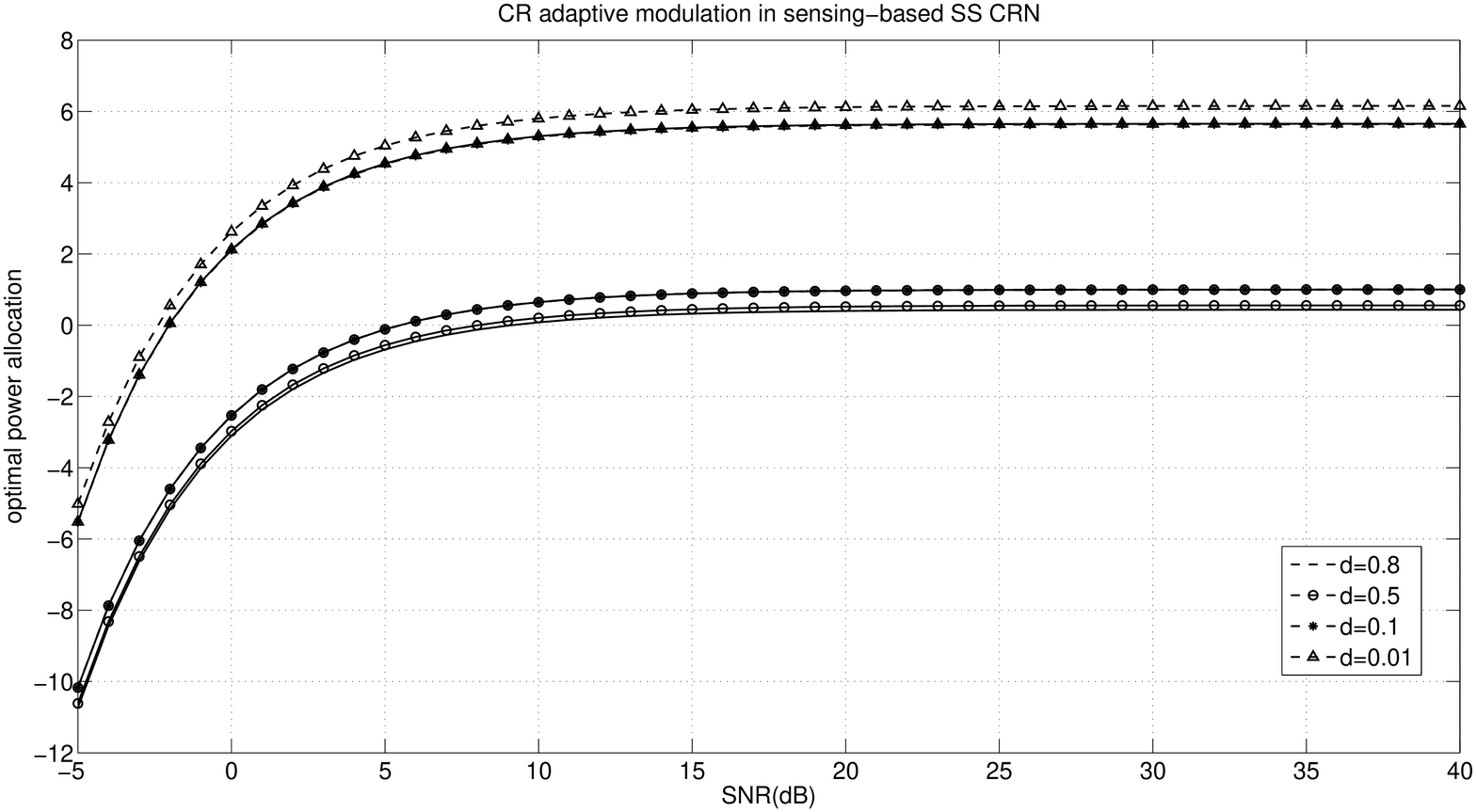}
\caption{Optimal power allocation $ P/ \overline{P} $  of CR adaptive modulation in sensing-based SS CRN for an average SNR $ \overline{\gamma}=10dB $ for different probabilities of detection $ d $  with peak interference power constraint $ I_{pk}=0dB $ (dashed lines) and $ I_{pk}=10dB $  }
\label{OPA_CR_10DB}
\end{figure}

\addtocounter{figure}{-1}
\addtocounter{subfigure4}{1}

\begin{figure}[t]
\centering
\includegraphics[width=5.5in, height=3.5in]{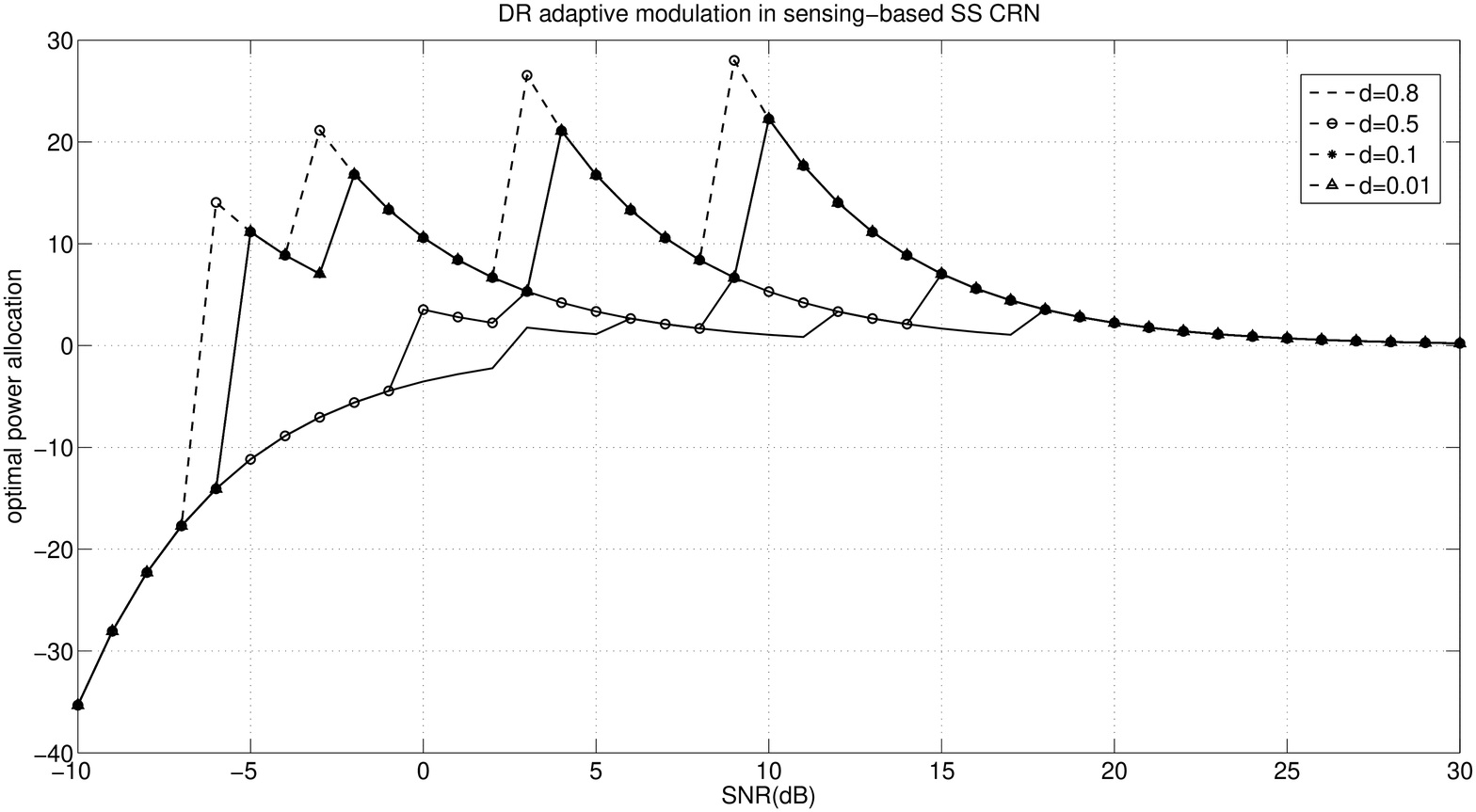}
\caption {Optimal power allocation $ P/ \overline{P} $  of DR adaptive modulation in sensing-based SS CRN for an average SNR $ \overline{\gamma}=10dB $  for different probabilities of detection $ d $  with peak interference power constraint $ I_{pk}=0dB $ (dashed lines) and $ I_{pk}=10dB $  }
\label{OPA_DR_10DB}
\end{figure}

\renewcommand{\thefigure}{\arabic{figure}}

\section{Conclusions}
In this work, the incorporation of adaptive modulation in CRNs over fading channels is studied thoroughly for the three main CRN types that have been identified so far i.e. OSA, SS and sensing-based spectrum sharing. We assume both CR and DR adaptive modulation schemes and we carry out the corresponding performance analysis. Closed-form expressions regarding the average spectral efficiency and the optimal power allocation for fading channels are derived incorporating the operational characteristics of both adaptive modulation and cognitive radio technologies. Numerical results for different numbers of employed fading regions and imposed interference power constraints are provided that reveal the performance gains regarding the achievable average spectral efficiency of adaptive modulation deployment in CRNs. Finally, detailed results that illustrate the optimal power allocation when an adaptive modulation scheme is considered for various target BER values and number of fading regions are provided. 
  
\section*{Acknowledgement}
This research work was supported by the Greek Government and the European Union in the context of the Archimedes III initiative.

\section*{Short CV}

\textbf{Fotis Foukalas} received the Diploma in Electrical and Computer Engineering from the Aristotle University of Thessaloniki, the M.Sc. degree in Communication Technologies from the  Electrical and Computer Engineering of National Technical University of Athens, and a Ph.D. in Wireless Mobile Communications from the Department of Informatics and Telecommunications, University of Athens. His areas of expertise are: cognitive radio, cross-layer optimization. 

\textbf{George T. Karetsos} received his diploma in electrical and computer engineering in 1992 and his Ph.D. in telecommunication systems in 1996, both from the National Technical University of Athens, Greece. He is currently an associate professor in the Information Technology and Telecommunications Department of the Technological Educational Institute of Larissa, Greece. His research interests are in the areas of wireless heterogeneous networking, performance evaluation and resource management for fixed and wireless networks.

\end{document}